\documentclass[aps,preprint]{revtex4}%
\usepackage{amsfonts}
\usepackage{amsmath}
\usepackage{amssymb}
\usepackage{graphicx}%
\begin{document}
\title{Analytic Treatment of Kapitza-Dirac Effect: Connecting Raman-Nath and Bragg Approximations}
\author{Gevorg Muradyan}
\author{Atom Zh. Muradyan}
\affiliation{Department of Physics, Yerevan State University 375025, Yerevan Armenia}
\keywords{Kapitza -Dirac diffraction, Bragg diffraction, Raman-Nath diffraction}
\pacs{03.75.-b}

\begin{abstract}
We develop an analytical approach for probability amplitudes of Kapitza-Dirac
effect that merge together the Raman-Nath and Bragg regimes of interaction.

\end{abstract}
\volumeyear{year}
\volumenumber{number}
\issuenumber{number}
\date[Date text]{date}
\received[Received text]{date}

\revised[Revised text]{date}

\accepted[Accepted text]{date}

\published[Published text]{date}

\startpage{1}
\maketitle
\tableofcontents

The Kapitza-Dirac effect constitutes diffraction of a structureless particle
(electron) by a standing electromagnetic wave, exhibiting the particle-wave
dual nature of matter in one of most convenient ways \cite{1}. It appears as a
counterpart to familiar optical diffraction by the periodic material grating.
Matter wave of the particle beam plays the role of the incoming wave while the
spatial periodicity of the grating interaction is ensured by the periodic
structure of the optical standing wave potential. Since its prediction in
1933, this phenomenon was addressed theoretically many times \cite{2,3} and
has been nicely observed experimentally by H. Batelaan et. al at
Nebraska-Lincoln University \cite{4}. (Detailed content can be found in review
article \cite{5} and dissertation \cite{6}).

\qquad In frame of 1D model of sinusoidal periodic potential the electron wave
function has a form%
\begin{equation}
\Psi\left(  z,t\right)  =\sum_{n=-\infty}^{+\infty}c_{n}(t)\exp\left[
i\left(  n_{0}+n\right)  2kz\right]  \tag{1}\label{1}%
\end{equation}
and the problem reduces to the following difference-differential equation for
$c_{n}\left(  t\right)  $ probability amplitudes of the diffraction
modes\cite{7}:%
\begin{equation}
\left(  \frac{d}{dt}+i\omega_{r}\left(  n_{0}+n\right)  ^{2}+iU_{0}\right)
c_{n}\left(  t\right)  =-i\frac{U_{0}}{2}\left(  c_{n-1}\left(  t\right)
+c_{n+1}\left(  t\right)  \right)  , \tag{2}\label{2}%
\end{equation}
$n=0,\pm1\pm2,...$. \ Here $k$ is the wave number of the running waves which
constitute the periodic potential, $n_{0}=p_{initial}/2\hbar k$ stands for the
normalized electron initial momentum, $\omega_{r}=E_{r}/\hbar$ with
$E_{r}=\left(  2\hbar k\right)  ^{2}/2m$ is the so called recoil frequency and
$U_{0}$ represents the amplitude of the ponderomotive potential in $\hbar$
units. \ In representation (1) the momentum transfer occurs in discrete units
$2\hbar k$ and generally is interpreted as absorption/stimulated emission of
photon pairs from the counterpropagating travelling waves, which give the
standing wave.

To clearly identify the limiting Raman-Nath and Bragg regimes it is convenient
to introduce a new amplitude%
\begin{equation}
C_{n}\left(  t\right)  =i^{n}\exp\left[  iU_{0}t+i\omega_{r}\left(
n_{0}+n\right)  ^{2}t\right]  c_{n}\left(  t\right)  \text{,} \tag{3}\label{3}%
\end{equation}
which transforms Eq.(2) to%
\begin{equation}
\frac{d}{dt}C_{n}\left(  t\right)  =\frac{U_{0}}{2}\left(  e^{i\omega
_{r}\left(  2n_{0}+2n-1\right)  t}C_{n-1}(t)-e^{-i\omega_{r}\left(
2n_{0}+2n+1\right)  t}C_{n+1}(t)\right)  \text{.} \tag{4}\label{4}%
\end{equation}
The analytic solution, presenting the Raman-Nath regime of interaction,
corresponds to the limiting case \ $\omega_{r}\left(  2n_{0}+2n-1\right)
t<<2\pi$, when the system (4) loses the time-dependent exponential
coefficients, transforming into equation with first kind Bessel function
solution: $C_{n}\left(  t\right)  =J_{n}(U_{0}t)$. Bessel function population
flow of diffraction modes has a dominantly double-peaked pattern,
symmetrically distributed about the initial state . This regime of realization
assumes that the change of the electron position along the standing wave
direction changes negligibly compared to the standing wave spatial period.

The second, Bragg regime of interaction distinguishes only discrete,
equidistant values or the electron initial momentum, namely in our notations
$n_{0}=\pm1/2\pm1,\pm3/2...$. Taking, for example, condition for the first
order diffraction $n_{0}=-1/2$ ($p_{0}=-\hbar k$), one can easily see that the
two amplitudes at the right hand side of system (4), \ $C_{0}(t)$ and
$C_{1}(t)$, lose the time dependence in exponential coefficients, while the
other ones preserve it. Assuming now an additional condition that $\omega
_{r}\left(  2n_{0}+2n-1\right)  t>>2\pi$ for any $\omega_{r}$ and $n$ (except,
of course, $n=0,1$), we get rapidly oscillating coefficients for terms
$n\neq0,1$ and thus almost totally suppress their contribution to the final
result (it reminds the rotating wave approximation widely used in the theory
of matter-laser resonance interactions). After neglecting all these terms, one
arrives to a simple pair of equations%
\begin{equation}
\frac{d}{dt}C_{0}(t)=-\frac{U_{0}}{2}C_{1}(t),\text{ }\frac{d}{dt}%
C_{1}(t)=\frac{U_{0}}{2}C_{0}(t)\text{,} \tag{5}\label{5}%
\end{equation}
resulting in $C_{0}(t)=\cos\left(  U_{0}t/2\right)  $ and $C_{1}%
(t)=\sin\left(  U_{0}t/2\right)  $ probability amplitudes for direct ($n=0$)
and Bragg diffraction ($n=1$).

There is no exact analytical solution to Kapitza-Dirac problem in frame of
Schr\H{o}dinger equation, which will be valid for any interaction time periods
and free of strict limitations on the system parameters. Diffraction
regularities have analytically been treated in mentioned regimes of
interaction and in close neighborhoods. They favor short- and long-time
regimes respectively and are also known as the thin- and thick-crystal approximations.

In this paper we develop a theory which treats the quantum particle (electron)
diffraction in the 1D periodic potential for any times of interaction. Our
formula quantitatively correctly describes both Raman-Nath and Bragg regimes
of interaction, thereby merging them into one essence of diffraction process. \ 

Our approach to the infinite system of equations (2) originates from the
remark that it connects the seeking amplitudes with opposite parity on the
left hand and right hand sides of the equations. Initial condition
$c_{n}(0)=\delta_{n,0}$ is, however, different for these two families of
amplitudes: all odd ones are zeroes, while the only nonzero member sits in the
even- manifold. Any approximate solution should be sensitive to the initial
conditions too, as we get rid of one of the parities in the set of equations
by introducing new, phase-shifted amplitudes%
\begin{equation}
\overline{c_{n}}(t)=i^{n}\exp[iU_{0}t]c_{n}(t) \tag{6}\label{6}%
\end{equation}
and arriving to%
\begin{align}
&
\begin{array}
[c]{c}%
\left(  \frac{d}{dt}+i\omega_{r}\left(  \left(  n_{0}+n\right)  ^{2}+1\right)
\right)  ^{2}\left(  \frac{d}{dt}+i\omega_{r}\left(  n_{0}+n\right)
^{2}\right)  \overline{c_{n}}(t)+\\
\left(  4\omega_{r}^{2}\left(  n_{0}+n\right)  ^{2}\left(  \frac{d}%
{dt}+i\omega_{r}\left(  n_{0}+n\right)  ^{2}\right)  +\frac{U_{0}^{2}}%
{2}\left(  \frac{d}{dt}+i\omega_{r}\left(  \left(  n_{0}+n\right)
^{2}+1\right)  \right)  \right)  \overline{c_{n}}(t)
\end{array}
\tag{7}\label{7}\\
&  =\frac{U_{0}^{2}}{2}\left(  \frac{d}{dt}+i\omega_{r}\left(  \left(
n_{0}+n\right)  ^{2}+1\right)  ^{2}\right)  \left(  \overline{c_{n-2}%
}(t)+\overline{c_{n+2}}(t)\right)  .\nonumber
\end{align}
It preserves the original tridiogonal form of (2) but is now a third order
differential - difference equation. In the following treatment the set (7)
will be regarded to describe the even- manifold of amplitudes.

We look for a trial solution of a definite integral form%
\begin{equation}
\overline{c_{n}}(t)=\exp\left[  -i\omega_{r}\left(  n_{0}+n\right)
^{2}t\right]  \frac{(-i)^{n}}{\pi}\int_{0}^{\pi}\cos\left(  n\varphi\right)
\exp\left[  i\lambda_{n}(\varphi)t\right]  d\varphi\text{,} \tag{8}\label{8}%
\end{equation}
where the $n$-dependent function $\lambda_{n}(\varphi)$ has to be determined
yet. Inserting Eq.(8r into (7) we obtain the following \ three difference
algebraic equations:%
\begin{equation}
\lambda_{n-2}(\varphi)+4\omega_{r}\left(  n_{0}+n-1\right)  -\lambda
_{n}(\varphi)=0 \tag{9}\label{9}%
\end{equation}%
\begin{equation}
\lambda_{n+2}(\varphi)+4\omega_{r}\left(  n_{0}+n+1\right)  -\lambda
_{n}(\varphi)=0 \tag{10}\label{10}%
\end{equation}%
\begin{align}
&  \left(  \left(  \lambda_{n}(\varphi)+\omega_{r}\right)  ^{2}\lambda
_{n}(\varphi)-4\omega_{r}^{2}\left(  n_{0}+n\right)  ^{2}\lambda_{n}%
(\varphi)-\frac{U_{0}^{2}}{2}\left(  \lambda_{n}(\varphi)+\omega_{r}\right)
\right)  \cos\left(  n\varphi\right) \tag{11}\label{11}\\
&  =\frac{U_{0}^{2}}{4}\left(  \lambda_{n-2}(\varphi)+\omega_{r}\left(
6n_{0}+6n-3\right)  \right)  \cos\left(  (n-2)\varphi\right)  +\frac{U_{0}%
^{2}}{4}\left(  \lambda_{n+2}(\varphi)-\omega_{r}\left(  2n_{0}+2n-3\right)
\right)  \cos\left(  (n+2)\varphi\right)  \text{.}\nonumber
\end{align}
Hence,for the trial function to be an exact solution of Eq.(7), the Eqs.
(9)-(11) should be identical to each other. Eq. (9) and (10) are really
mutually equivalent and one of them, for instance Eq. (9), can be put out from
consideration. The last one, Eq. (11), however, is not equivalent to Eqs. (9)
and (10). This means that the analytic form (8) can not be an exact solution
to the problem. Our approximation lies just in this point and amounts to
taking of Eqs. (9)-(11) as a system of three equations relative to
$\lambda_{n+2}(\varphi)$, $\lambda_{n-2}(\varphi)$ and seeking $\lambda
_{n}(\varphi)$. Then finding $\lambda_{n+2}(\varphi)$ as a linear function of
$\lambda_{n}(\varphi)$ and inserting it into Eq. (11), we arrive to a third
order algebraic equation%
\begin{equation}
\left(  \lambda_{n}(\varphi)+2\omega_{r}/3\right)  ^{3}+p_{n}\left(
\varphi\right)  \left(  \lambda_{n}(\varphi)+2\omega_{r}/3\right)
+q_{n}\left(  \varphi\right)  =0 \tag{12}\label{12}%
\end{equation}
with real coefficients%
\[
p_{n}\left(  \varphi\right)  =-\left(  U_{0}^{2}\cos^{2}\left(  \varphi
\right)  +4\omega_{r}^{2}\left(  n_{0}+n\right)  ^{2}+\omega_{r}^{2}/3\right)
\]
and%
\[
q_{n}\left(  \varphi\right)  =-\frac{2}{27}\omega_{r}^{3}-U_{0}^{2}\omega
_{r}\left(  \frac{1}{3}\cos^{2}\left(  \varphi\right)  +\left(  n_{0}%
+n\right)  \cos\left(  2\varphi\right)  \right)  +\frac{8}{3}\omega_{r}%
^{3}\left(  n_{0}+n\right)  ^{2}\text{.}%
\]
The sign of the first coefficient is always negative and depending on the sign
of quantity
\[
Q_{n}\left(  \varphi\right)  =\left(  \frac{p_{n}\left(  \varphi\right)  }%
{3}\right)  ^{3}+\left(  \frac{q_{n}\left(  \varphi\right)  }{2}\right)  ^{2}%
\]
one has two distinct forms for the solutions of Eq. (12)\cite{8}.

Later on we'll denote the three roots of Eq. (12) as $\lambda_{n}(j;\varphi)$,
$j=1,2,3$, the corresponding amplitudes as $\overline{c_{n}}(j;t)$ and the
respective wave functions as $\Psi\left(  j;z,t\right)  $. Then the general
form of probability amplitudes should be written as%
\begin{equation}
\Psi\left(  z,t\right)  =h_{1}\Psi\left(  1;z,t\right)  +h_{2}\Psi\left(
2;z,t\right)  +h_{3}\Psi\left(  3;z,t\right)  \tag{13}\label{13}%
\end{equation}
or, equivalently%
\begin{align}
c_{n}(t)  &  =\exp\left[  -i\omega_{r}\left(  n_{0}+n\right)  ^{2}%
t-iU_{0}t\right]  \frac{\left(  -1\right)  ^{n}}{\pi}\times\tag{14}%
\label{14}\\
&  \int_{0}^{\pi}\left(  h_{1}\exp\left[  i\lambda_{n}(1;\varphi)t\right]
+h_{2}\exp\left[  i\lambda_{n}(2;\varphi)t\right]  +h_{3}\exp\left[
i\lambda_{n}(3;\varphi)t\right]  \right)  \cos\left(  n\varphi\right)
d\varphi\text{.}\nonumber
\end{align}
Three $h$-coefficients are determined from initial conditions for the wave
function (13) and its first and second derivatives:%
\[
h_{1}+h_{2}+h_{3}=1,
\]%
\begin{equation}
l_{1}h_{1}+l_{2}h_{2}+l_{3}h_{3}=0, \tag{15}\label{15}%
\end{equation}%
\[
L_{1}h_{1}+L_{2}h_{2}+L_{3}h_{3}=U_{0}^{2}/2,
\]
with the following notations for coefficients:%
\begin{equation}
l_{j}=\frac{1}{\pi}\int_{0}^{\pi}\lambda_{n=0}\left(  j;\varphi\right)
d\varphi,\text{ \ \ }L_{j}=\frac{1}{\pi}\int_{0}^{\pi}\lambda_{n=0}\left(
j;\varphi\right)  ^{2}d\varphi\text{,} \tag{16}\label{16}%
\end{equation}
$j=1,2,3$. \ This step crowns the procedure and hence the formula (14)
presents the solution of the problem for any even $n$, the amount of acquired
momentum in $2\hbar k$ units .

To determine still untouched odd- probability amplitudes, we have to return to
original equation (1) and shift the numbering by one:%
\begin{equation}
\left(  \frac{d}{dt}+i\omega_{r}\left(  n_{0}+n\right)  ^{2}+iU_{0}\right)
c_{n+1}\left(  t\right)  =-i\frac{U_{0}}{2}\left(  c_{n}\left(  t\right)
+c_{n+2}\left(  t\right)  \right)  , \tag{17}\label{17}%
\end{equation}
Inserting even-$n$ solutions into the right hand side of (17) and simply
integrating equation with zero initial condition, we complete our approach to
analytic solution of the stated Kapitza-Dirac diffraction problem.%
\begin{figure}
[ptb]
\begin{center}
\includegraphics[
height=2.6282in,
width=3.9332in
]%
{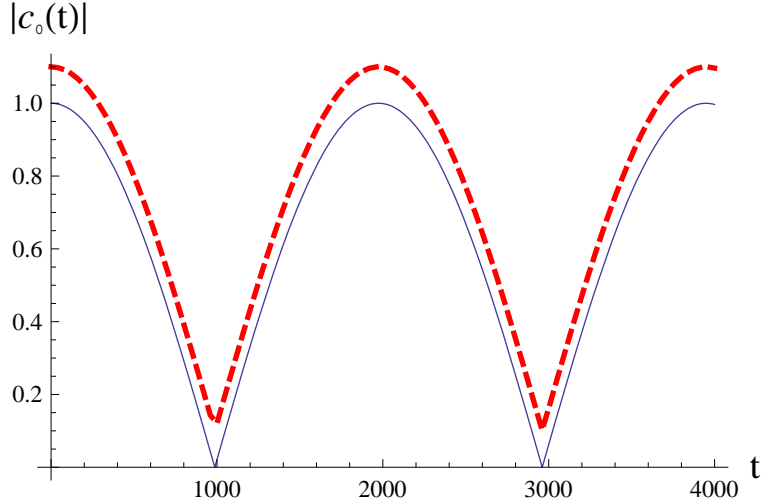}%
\caption{(Color online) $n=0$ diffraction probability amplitude evalution in
Bragg diffraction regime. Solid line gives the result of exact numerical
simulation. \ Dashed line is the amplitude evolution for analytic formula
described in the text. The dashed line is shifted vertically by $0.1$ to make
the two lines visually different(we will make this shift in all figures) .
Horizaontal time axis is normalized to green light ($\lambda_{green}%
=6000\mathring{A}$) Compton backscattering frequency shift $\Delta
\omega_{Compton}=4\pi\hbar\omega_{green}^{2}/M_{e}c^{2},$ where $M_{e}$ is the
elelctron mass. Other parameters are $n_{0}=-0.5$, $U_{0}=0.01\Delta
\omega_{Compton}/\pi\,$, $\omega_{r}=\Delta\omega_{Compton}/\pi$.}%
\end{center}
\end{figure}
\begin{figure}
[ptb]
\begin{center}
\includegraphics[
height=2.6282in,
width=3.9332in
]%
{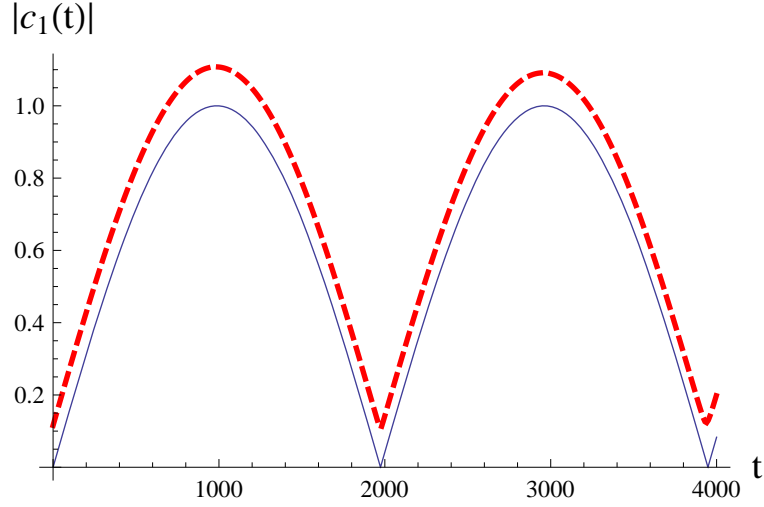}%
\caption{(Color online) $n=1$ diffraction probability amplitude evalution in
Bragg diffraction regime. All the parameters are as in Fig.1.}%
\end{center}
\end{figure}
\begin{figure}
[ptb]
\begin{center}
\includegraphics[
height=2.6429in,
width=3.9332in
]%
{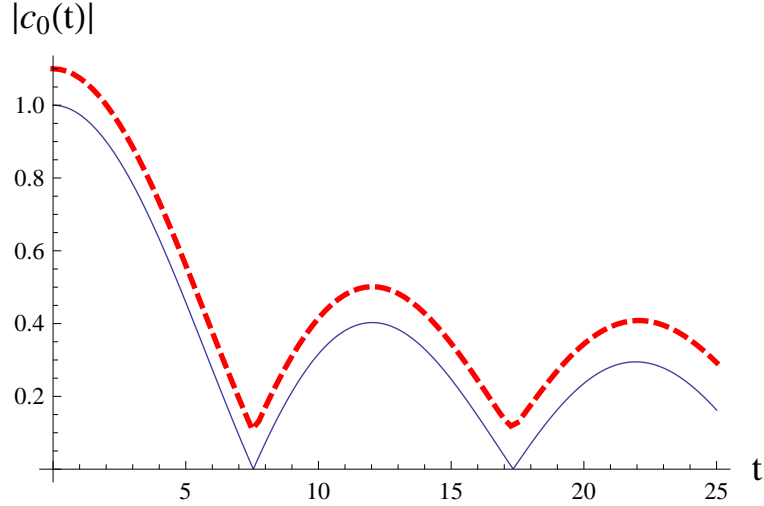}%
\caption{(Color online) $n=0$ diffraction probability amplitude evalution in
Raman-Nath diffraction regime. $n_{0}=-0.5$, $U_{0}=\Delta\omega_{Compton}%
/\pi\,$, $\omega_{r}=0.001\Delta\omega_{Compton}/\pi$.}%
\end{center}
\end{figure}
\begin{figure}
[ptb]
\begin{center}
\includegraphics[
height=2.6152in,
width=3.9332in
]%
{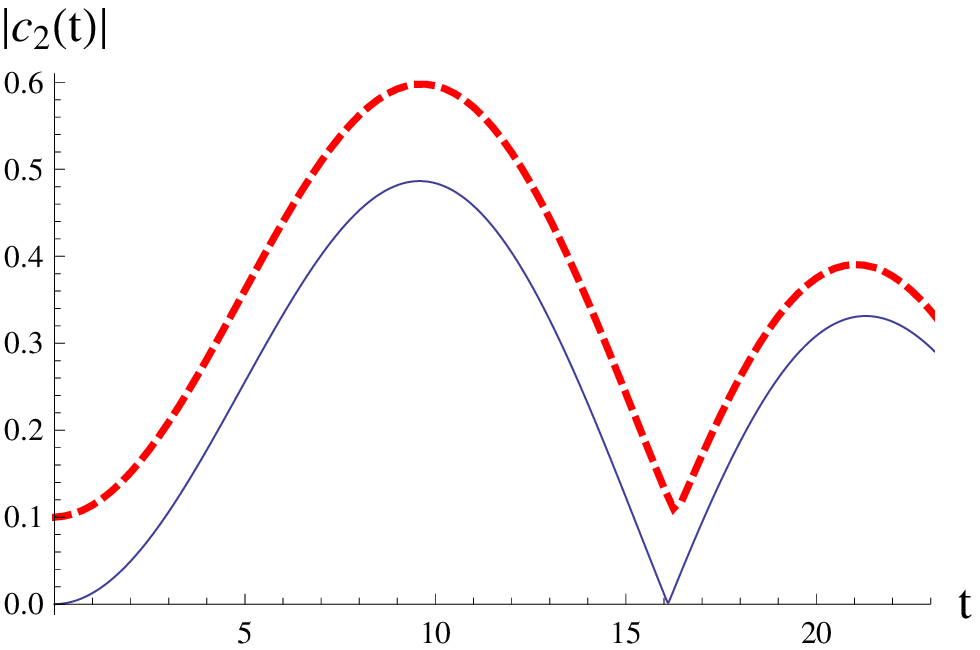}%
\caption{(Color online) $n=2$ diffraction probability amplitude evalution in
Raman-Nath diffraction regime. All the parameters are as in Fig.3.}%
\end{center}
\end{figure}

To value the developed analytic approximation, we have compared its results
with the exact numerical solutions of the original set of Eqs. (2). In order
to implement these simulations we have used the Crank-Nicolson method
\cite{9}. Comparison shows that the presented approximation works excellent in
both, Raman-Nath and Bragg regimes of interaction. Two particular cases are
illustrated in Figs.1-4. The graphs in each figure are indistinguishable from
each other at sight and are shifted in vertical direction in illustrative purposes.

In intermediate regimes (relative to Raman-Nath and Bragg) our numerical
calculation has definite restrictions, connected with the limitations on
calculation of inverse matrices required by the Crank-Nicolson method. \ Thus
ultimate conclusions here cant be done yet. \ However, every case when we were
sure in the correctness of numerical calculations the coincidence between our
approximate \ analytical results and numerical ones was quiet good. \ Fig. 5
and 6 illustrate such a case with parameter values $U_{0}\approx\omega_{r}$
(note that the Raman-Nath approximation requires $U_{0}>>\omega_{r},$ and the
Bragg approximation - the opposite one $U_{0}<<\omega_{r}$ ).%
\begin{figure}
[ptb]
\begin{center}
\includegraphics[
height=2.6429in,
width=3.9332in
]%
{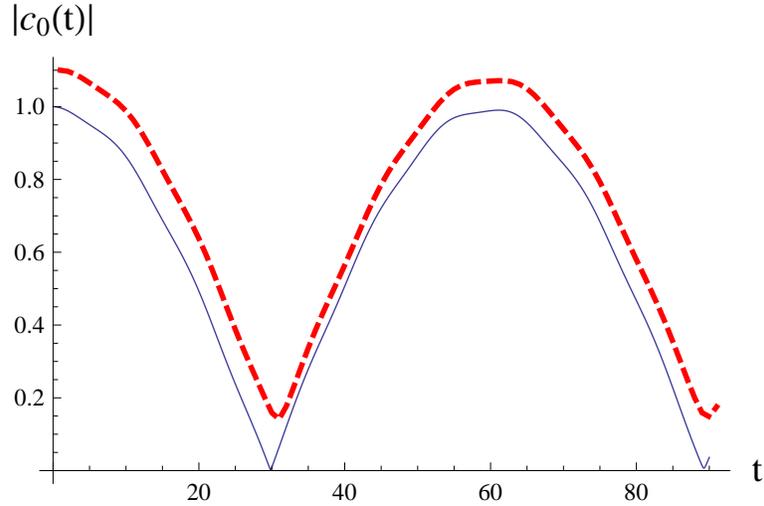}%
\caption{(Color online) $n=0$ diffraction probability amplitude evalution in
intermediate diffraction regime. $n_{0}=-0.5$, $U_{0}=\Delta\omega
_{Compton}/3\pi\,$, $\omega_{r}=\Delta\omega_{Compton}/\pi$.}%
\end{center}
\end{figure}
\begin{figure}
[ptb]
\begin{center}
\includegraphics[
height=2.6282in,
width=3.9332in
]%
{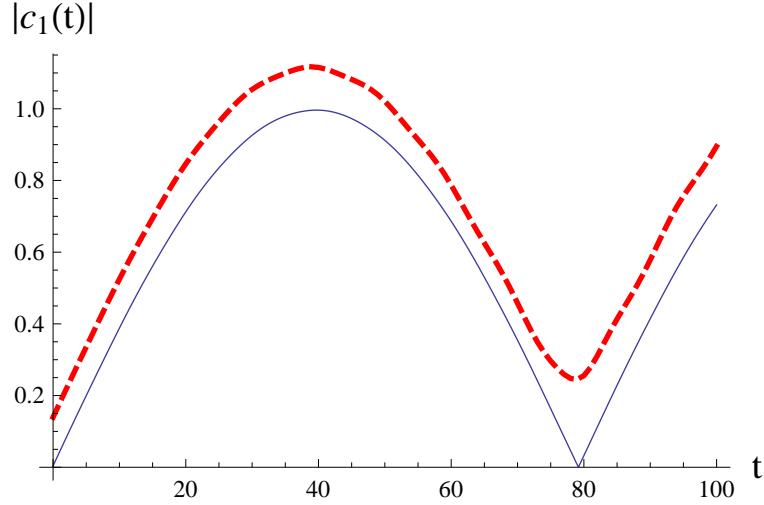}%
\caption{(Color online) $n=1$ diffraction probability amplitude evalution in
intermediate diffraction regime. $n_{0}=-0.5$, $U_{0}=\Delta\omega
_{Compton}/4\pi\,$, $\omega_{r}=\Delta\omega_{Compton}/\pi$.}%
\end{center}
\end{figure}

This gives some credibility to the presented analytical approximation over the
intermediate range of parameters too and in the future we will endeavour in
reaching a full definiteness in this direction too.

This work is supported by Alexander von Humboldt foundation and NFSAT/CRDF
Grant No. UCEP-0702.

\end{document}